\documentclass[11pt, A4]{article}

\usepackage{amsfonts}

\usepackage{cite}
\usepackage{graphicx}
\usepackage{epstopdf}
\usepackage{amsmath}
\usepackage{amssymb}
\usepackage{amsthm}
\usepackage[usenames]{color}
\usepackage{array}

\usepackage[
      colorlinks=true,
      linkcolor=blue,
      urlcolor=blue,
      filecolor=blue,
      citecolor=red,
      pdfstartview=FitV,
      pdftitle={},
      pdfauthor={},
      pdfsubject={},
      pdfkeywords={},
      pdfpagemode=None,
      bookmarksopen=true
]{hyperref}

\usepackage{epsfig}
\usepackage{mathtools}

\usepackage{hyperref}

\newcommand{\mysection}[1]{\section{#1}\setcounter{equation}{0}}

\def\d{\partial}
\def\e{\epsilon}

\def\be{\begin{equation}}
\def\ee{\end{equation}}
\def\bea{\begin{eqnarray}}
\def\eea{\end{eqnarray}}

\newcommand{\rom}[1]{\mathrm{#1}}

  \usepackage{comment}

\begin{document}
 \begin{centering}
  
{\Large  {\bf
Supersymmetric Black Hole Hair and $AdS_3 \times S^3$} } \\

 \vspace{0.8cm}

 {Subhodip Bandyopadhyay$^{a,b}$, Yogesh K.~Srivastava$^{a,b}$, Amitabh Virmani$^c$}
\vspace{0.8cm}

\begin{minipage}{.9\textwidth}\small  \begin{center}
$^a${National Institute of Science Education and Research (NISER), \\ Bhubaneswar, P.O. Jatni, Khurda, Odisha, India 752050}\\
  \vspace{0.5cm}
$^b$Homi Bhabha National Institute, Training School Complex, \\ Anushakti Nagar, Mumbai, India 400085 \\
  \vspace{0.5cm}
$^c$Chennai Mathematical Institute, H1 SIPCOT IT Park, \\ Kelambakkam, Tamil Nadu, India 603103\\
  \vspace{0.5cm}
{\tt \{subhodip.bandyopadhyay, yogeshs\}@niser.ac.in, avirmani@cmi.ac.in}
\\ $ \, $ \\

\end{center}
\end{minipage}

\end{centering}

\begin{abstract}
The 4D-5D connection allows us to view the same near horizon geometry as part of a 4D black hole or a 5D  black hole. A much studied example of this phenomenon is the BMPV black hole uplifted to 6D with flat base space versus Taub-NUT base space. These black holes have identical near horizon $AdS_3 \times S^3$ geometry. In this paper, we  study modes in $AdS_3 \times S^3$ and
 identify those that correspond to supersymmetric  hair modes in the full black hole spacetimes. We show that 
these modes 
satisfy 
non-normalisable boundary conditions in $AdS_3$.
The non-normalisable boundary conditions are different for different hair modes and for different asymptotic completion. 
We also discuss how the supersymmetric hair modes on 
BMPV black holes fit into the classification of supersymmetric solutions of 6D supergravity. 
 
 \end{abstract}

\newpage

\tableofcontents

\setcounter{equation}{0}

\section{Introduction} 

String theory has been very successful in computing the index/degeneracy for a class of supersymmetric black holes \cite{Strominger:1996sh, 0708.1270}. In some cases, exact  counting formulas are known for helicity trace indices in terms of certain modular forms. Can these precise microscopic answers be reproduced  via a computation on the gravity side?

A concrete proposal for such a computation is to do a string path integral in the near horizon geometry of the black hole \cite{Sen:2008vm}. An year after this proposal was made, a puzzle regarding this was outlined along with a possible resolution in \cite{Banerjee:2009uk, Jatkar:2009yd}. The puzzle goes as follows: the Breckenridge-Myers-Peet-Vafa (BMPV) black hole \cite{Breckenridge:1996is} in flat space and in Taub-NUT space have identical near-horizon geometry but different microscopic indices. The same near-horizon geometry cannot account for different microscopic answers.

 It is now well appreciated that the difference is due to the presence of black hole hair modes: smooth, normalizable,  bosonic and fermionic degrees of freedom living outside the horizon. Some of these extra degrees of freedom were identified and constructed in \cite{Banerjee:2009uk, Jatkar:2009yd}. These computations were extended to a wide class of other models in \cite{Chakrabarti:2020ugm} where the need to include both  the twisted and untwisted sector hair modes was identified and many gaps in the earlier calculations were filled. See also \cite{Chattopadhyaya:2020yzl}.
 
Given these hair modes,  it is unclear how to match a path integral computation in the near horizon geometry to the exact microscopic formulas. On the one hand, the hair modes must be taken into account for the matching to be fully successful \cite{Sen:2009vz,Dabholkar:2010rm}. On the other hand, accounting for hair modes from the near-horizon perspective seems difficult.  Recent developments \cite{Iliesiu:2021are, Sen:2023dps} suggest that the path integral over modes living in full asymptotically flat space-time geometry might account for the near-horizon as well as the hair mode contributions in a unified approach. In view of these discussions, it is timely to study hair modes from the near-horizon perspective.  In this paper, we undertake such a study.

We  explore modes in $AdS_3 \times S^3$ and
 identify those that correspond to the known supersymmetric  hair modes  in the full black hole spacetimes.  A key motivation for this study is to understand these hair modes from the near-horizon perspective. Another motivation is to understand the relation between the hair deformations to other  deformations of $AdS_3$ such as null-warped $AdS_3$.  We explore in detail the boundary conditions the hair modes satisfy in $AdS_3$.   
In particular, we show that all these modes satisfy non-normalizable boundary conditions in $AdS_3$. From our analysis, it is clear that any computation on the gravity side in the near-horizon region that attempts to reproduce the precise microscopic answers requires the use of non-normalizable modes.

 The rest of the paper is organised as follows. 

In section \ref{sec:review}, we briefly review black hole hair modes identified in ref.~\cite{Jatkar:2009yd} as the Garfinkle-Vachaspati (GV) wave modes and as the anti-self-dual form field hair modes. 
  
In section \ref{sec:GV}, we study  GV wave modes in $AdS_3 \times S^3$.  Expanding the GV modes in spherical harmonics on the $S^3$, we show that the GV hair modes on the BMPV black hole in flat spacetime correspond to $L=1$ spherical harmonics while the GV hair modes on the Taub-NUT black hole correspond to  $L=2$ spherical harmonics.  We show that all these modes are
non-normalisable in $AdS_3$. Of course, when embedded in the full black hole geometry with flat space asymptotics, these modes are normalisable. We study  possible singularities of these modes in the $AdS_3 \times S^3$ near-horizon geometry and we show that the $L=2$ modes are regular at the horizon. The $L=1$ modes turn out to be singular at the horizon with certain components of the Riemann tensor diverging. This is exactly in line with the analyses of refs.\cite{Jatkar:2009yd, Chakrabarti:2020ugm}. 

At the linearised level, it was already understood in ref.~\cite{Banerjee:2009uk} that hair modes are located outside the near-horizon region. The analysis of this paper makes those observations more quantitative. It is useful to appreciate another context where a closely related issue is well explored, namely,  quasi-normal modes of microstate geometries \cite{Chowdhury:2007jx, Chakrabarty:2015foa, Chakrabarty:2019ujg,Bena:2020yii}. In this literature, it is well appreciated that in the limit when a microstate geometry has a  large 
$AdS_3 \times S^3$ region, 
quasi-normal modes correspond to excitations  that leak out from the near-core $AdS_3 \times S^3$ region to flat space region. The leaking excitations satisfy non-normalisable boundary condition at the edge of $AdS_3$ (otherwise they won't leak out).  Our results are exactly the same, but in the context of the supersymmetric hair modes. The hair mode ``leak out'' from the near-throat region. The hair modes are supported outside the throat region and they decay to zero in the asymptotic region.

In section \ref{sec:form_field}, we study the deformation due to turning on an anti-self-dual three-form field strength in $AdS_3 \times S^3$.  Using the decomposition of the anti-self-duality equation in spherical harmonics, following~\cite{Maldacena:1998bw}, we derive equations for such perturbations in $AdS_3$. Solving these equations, we identify the anti-self-dual field strength that correspond to a hair mode of  the original black hole in Taub-NUT space. Then, solving the Einstein equations in $AdS_3 \times S^3$ with this anti-self-dual field as an additional source, we obtain the deformed metric. We do a regularity  analysis for this mode, and find that both the metric and the anti-self dual field strength are regular at the $AdS_3$ horizon. As for the case of the GV modes, the anti-self-dual field strength deformations are also not asymptotically $AdS_3 \times S^3$ but correspond to non-normalizable deformations even though in the full black hole geometry they are normalizable.   Moreover, we find that the 
non-normalizable metric deformation in this case is closely related to null-warped $AdS_3$.

In section \ref{sec:fermionic}, we make similar observations on the non-normalisability of the fermionic hair modes of refs.~\cite{Jatkar:2009yd, Chakrabarti:2020ugm}.  We close with a brief discussion in section \ref{sec:discussion}. In appendix \ref{app:6D_sugra}, we embed the hair modes of the full   black holes in minimal six-dimensional supergravity coupled to a tensor multiplet.

\section{Review: black holes, near horizon geometry, and bosonic hair modes} 
\label{sec:review}

In this section, we review the construction of bosonic hair modes for  $J_L=0$ BMPV black hole in flat space and Taub-NUT space \cite{Banerjee:2009uk, Jatkar:2009yd, Chakrabarti:2020ugm}.

\subsection*{Black hole in flat space}
The $J_L=0$ BMPV black hole \cite{Breckenridge:1996is} metric uplifted to six-dimensions takes the form,
\bea
ds^2 &=& G_{MN} dx^M dx^N \\
&=& \psi^{-1}(r) \left[ du \, dv + (\psi(r)-1)dv^2  \right] + \psi(r) ds^2_\rom{flat},
\eea
where
\be
u = x^5 -t, \qquad \mbox{and} \qquad v =  x^5 + t. 
\ee
Here $x^5$ is the coordinate on the sixth dimension, a Kaluza-Klein circle denoted $S^1$ with period $2 \pi R_5$. The function $\psi(r)$ appearing in the metric is
\be
\psi(r) = 1 + \frac{r_0}{r}, \label{psi-r}
\ee
with $r_0$ related to the charges carried by the black hole. The coordinate  $r$ is the Gibbons-Hawking radial coordinate on 4D flat space introduced as follows. The spherical polar coordinates for four-dimensional Euclidean flat space $(\tilde r, \tilde \theta, \tilde \phi,\tilde \psi)$, in which the metric takes the form, 
\be
ds^2_\rom{flat} = d \tilde r^2 + \tilde r^2 (d\tilde \theta^2 + \cos^2\tilde \theta d \tilde \phi^2 +  \sin^2\tilde \theta d \tilde \psi^2),
\ee
are related to cartesian coordinates as, 
\begin{align}
w^1 &= \tilde r \cos \tilde \theta \cos \tilde \phi, &
w^2 &= \tilde r \cos \tilde \theta \sin \tilde \phi,\\
w^3 &= \tilde r \sin \tilde \theta \cos \tilde \psi ,&
w^4 &= \tilde r \sin \tilde \theta \sin \tilde \psi. \label{w-coords}
\end{align}
The Gibbons-Hawking coordinates $(r,\theta, \phi, x^4)$ are defined via,
\begin{align}
\tilde r &= 2 \sqrt{r}, & \label{GH-r}
\tilde \theta &= \frac{\theta}{2},\\
 \tilde \phi &= \frac{1}{2} (x^4 + \phi), &
 \tilde \psi &= \frac{1}{2} (x^4 - \phi). 
 \end{align}
In these coordinates flat space metric takes the form
\be
ds^2_\rom{flat} = \frac{1}{r}dr^2 + r (d\theta^2 + \sin^2 \theta d\phi^2 + (dx^4+\cos \theta d\phi)^2). \label{GH-flat}
\ee
For details on the identifications of the angular coordinates, we refer the reader to refs.~\cite{Jatkar:2009yd, Chakrabarti:2020ugm}.

The other fields supporting the solutions are as follows. The six-dimensional dilaton is set to its constant asymptotic value throughout the spacetime,
\be
e^\Phi = \lambda.
\ee 
The only other nontrivial field is the self-dual three-form Ramond-Ramond (RR) field $F^{(3)}$. It takes the form
\be
F^{(3)} = \frac{r_0}{\lambda} \left(\epsilon_3 + \star_6 \epsilon_3 \right), \label{F3}
\ee
where  
$\epsilon_3 = \sin \theta \ dx^4 \wedge d\theta  \wedge d\phi $
and $\epsilon^{t54r\theta\phi} = +1$.

\subsection*{Black hole in Taub-NUT space}

Next we consider $J_L = 0$ BMPV black hole in Taub-NUT space \cite{Gauntlett:2002nw, hep-th/0503217}, following the notation of refs.~\cite{Jatkar:2009yd, Chakrabarti:2020ugm}.
The metric of the four dimensional Taub-NUT space in Gibbons-Hawking coordinates is given  as a U(1) fibre over an $\mathbb{R}^3$ base space,
\be
ds^2_{TN} = \left( \frac{4}{R_4^2} + \frac{1}{r} \right)^{-1} (dx^4 + \cos \theta d\phi)^2 + \left(\frac{4}{R_4^2} + \frac{1}{r}\right) (dr^2 + r^2 d \theta^2 + r^2 \sin^2 \theta d\phi^2).
\label{metric-TN}
\ee
Compared to flat space in Gibbons-Hawking coordinates written above, the only difference is that the $\frac{1}{r}$ factors are now replaced with $\left( \frac{4}{R_4^2} + \frac{1}{r} \right)$.   The $x^4$ coordinate labels the circle ${\widetilde S}^1$ and it is periodic with size $2 \pi R_4$ at infinity. 

The  metric describing the $J_L =0$ BMPV black hole in Taub-NUT space takes the form, 
\be
\label{metric_TN}
ds^2 = \psi^{-1}(r) [du \, dv + (\psi(r) -1) dv^2 ] + \psi(r) ds^2_{TN}.  
\ee
The dilaton is set to its asymptotic value $e^{\Phi} = \lambda$.
The self-dual field strength $F^{(3)}$ supporting this solution is,
\be
 F^{(3)} = \frac{r_0}{\lambda}\Big[ (\e_3 + \star_6 \e_3)  \Big].
 \label{F3_TN}
\ee

\subsection*{Near-horizon geometry}

It is well known that the two black holes discussed above have the identical near-horizon geometry \cite{hep-th/0505094, Banerjee:2009uk}. For the black hole in flat space, we can take the near-horizon limit by taking the limit $r \rightarrow 0$  
to get
\be
ds^2 = \frac{r}{r_0} \left(dudv + \frac{r_0}{r} dv^2 \right) + \frac{r_0}{r^2}dr^2 + r_0 \left( d\theta^2 + \sin^2\theta d\phi^2 + (dx^4 + \cos\theta d\phi)^2 \right).\label{nearhorizon_GH}
\ee
This metric describes extremal BTZ $\times~S^3$. To bring the metric in a more standard form,  we perform the following coordinate transformation
\be
\phi = \tilde{\phi} -\tilde{\psi} \ , \qquad  \theta = 2\tilde{\theta} \ , \qquad x^4 = \tilde{\phi} +\tilde{\psi} \ , \qquad  \frac{r}{r_0}= \frac{\ell^2}{z^2} , \qquad r_0 = \frac{1}{4}\ell^2,\label{FLct}
\ee
to get
\begin{equation}
ds^2= \frac{\ell^2}{z^2} \left(du \, dv  + dz^2 + \frac{z^2}{\ell^2} dv^2  \right)+ \ell^2 
d\Omega^2_{S^3} , \label{nearhorizon}
\end{equation}
with 
\be
d\Omega^2_{S^3} = d \tilde{\theta}^2 + \cos^2 \tilde{\theta} d \tilde{\phi}^2 + \sin^2 \tilde{\theta} d \tilde{\psi}^2.
\ee
Here, $\ell$ is the $AdS_3$ length. The boundary of $AdS_3$ is at $z=0$. The horizon of the extremal BTZ is at $z \to \infty$.

Metric \eqref{nearhorizon} can be simplified further by introducing Poincar\'e coordinates in the $AdS_3$ part of the geometry. 
The transformation 
\begin{align}
\bar v &= e^{\frac{2v}{\ell}}, &
\bar u &= u - \frac{1}{\ell} z^2, &
\bar z &=  \sqrt{\frac{2}{\ell}}  z e^{\frac{v}{\ell}}. \label{Poincare}
\end{align}
takes us to the Poincar\'e coordinates $(\bar u, \bar v, \bar z )$.\footnote{This coordinate transformation is a special case of Banados, Chamblin, and Gibbons \cite{Banados:1999tw} transformations. They are discussed in more detail in section \ref{sec:L_0}.}
The final metric takes the form
\begin{equation}
ds^2= \frac{\ell^2}{\bar z^2} \left(d \bar u \, d \bar v  + d\bar z^2 \right)+ \ell^2 
d\Omega^2_{S^3} . \label{nearhorizon_new}
\end{equation}
The periodic identifications of $u$ and $v$ coordinates are not so natural in the Poincar\'e coordinates $(\bar u, \bar v, \bar z )$. The horizon is located at $\bar z \to \infty$. For local analyses (such as regularity of the modes), Poincar\'e coordinates are the most convenient to work with. For identifying modes in $AdS_3$ with the hair modes of the black holes, $(u, v, z)$ are the most convenient coordinates to work with.

\subsection*{Bosonic hair modes on black hole in flat space}
Now we consider the hair modes~\cite{Jatkar:2009yd, Chakrabarti:2020ugm}. The solution-generating technique of Garfinkle and Vachaspati \cite{garfinkle:1990, myers:1997} was used to add hair modes to the above systems. Given a space-time metric $G_{MN}$ with null Killing vector $k_M$ that is hypersurface orthogonal, i.e.,  
\be
\nabla_{[M}k_{N]} = k_{[M}\nabla_{N]} A
\ee
for some function $A$,  new exact solutions to the supergravity equations are constructed by the following transform,
\be
G'_{MN} = G_{MN} + e^A \, T \, k_M \, k_N, \label{GV_main}
\ee
 and matter fields remain unchanged. The transformed metric is  a valid solution if $T$ satisfies,
\be
\Box T = 0  \qquad \: \text{and} \: \qquad k^M \d_M T = 0,
\ee
where $\Box$ is d'Alembertian with respect to the background metric $G_{MN}$. 

The BMPV black string in six-dimensions possesses such a  null Killing vector, 
\be
k^M \partial_M = \frac{\partial}{\partial u},
\ee 
with \be
e^A = \psi.
\ee Applying the Garfinkle-Vachaspati transform we get,
\be
ds^2 = \psi^{-1} \left[ du \, dv + (\psi -1 + T(v, \vec w)) \,dv^2  \right] + \psi \, ds^2_\rom{flat}. \label{deformed_BMPV}
\ee
Requiring regularity at infinity and at the origin and keeping only terms that cannot be removed by coordinate transformations \cite{Dabholkar:1995nc}, we can choose
\be
T(v, \vec w) = f_i(v) w^i, \qquad \qquad  \int_0^{2 \pi R_5} f_i(v) dv = 0,
\ee
with four arbitrary functions $f_i(v)$. The deformed metric \eqref{deformed_BMPV} does not look  asymptotically flat, but via a standard change of coordinates \cite{Dabholkar:1995nc} it can be brought to a manifestly asymptotically flat form. We can take the near-horizon limit of the deformed metric by taking the $r \rightarrow 0$ limit. It gives
\begin{equation}
ds^2= \frac{\ell^2}{z^2} \left(dudv  + dz^2 +  \left(\frac{z^2}{\ell^2} + T (z)\right) dv^2 \right)+ \ell^2 d\Omega^2_{S^3}.
\end{equation}
where 
\be
T(z)= f_i(v)w^i(z). 
\ee

\subsection*{Bosonic hair modes on black hole in Taub-NUT space}

Next we recall the hair mode deformations of  the BMPV black hole in Taub-NUT space. A class of these deformations is generated by the Garfinkle-Vachaspati transform.  The deformed metric takes the form
\be
ds^2 = \psi^{-1}(r) [du \, dv + (\psi(r) -1 + \widetilde{T}(v, x^4, r, \theta, \phi)) \, dv^2 ] + \psi(r) \, ds^2_{TN},
\ee
where now the condition is that $\widetilde{T}(v, x^4, r, \theta, \phi)$ is a harmonic function on four-dimensional Taub-NUT space. For an $x^4$ independent function, the condition simply reduces to the function $\widetilde{T}$ being harmonic on the three-dimensional transverse space $\mathbb{R}^3$ spanned by $(r, \theta, \phi)$. As in the BMPV case, requiring the deformation to be regular at the origin and at infinity and dropping terms that can be removed by  coordinate transformations, we can choose
\begin{equation}
    \widetilde{T}(v, \vec y)= g_i(v)y^i, \qquad \qquad \int_0^{2\pi R_5}g_i(v)dv=0, \label{gvTN}
\end{equation}
where $y^i$ are cartesian coordinates on $\mathbb{R}^3$ and $g_i(v)$ are three arbitrary functions.  As before, we can take the near-horizon limit of the deformed metric by taking the $r \rightarrow 0$ limit.

The BMPV black hole in Taub-NUT space admits another class of deformations \cite{Jatkar:2009yd} sourced by anti-self-dual three-forms.  To keep things as simple as possible, in this paper, we consider only one such three-form to be present.  These new deformations arise (essentially) because the Taub-NUT space admits a self-dual harmonic form,
\be
\omega_{TN} = - \frac{r}{4 r + R_4^2} \sin \theta \: d\theta \wedge d\phi + \frac{R_4^2}{(4r + R_4^2)^2} dr \wedge (dx^4 + \cos \theta \, d \phi), \label{omega_TN}
\ee
where $\epsilon_{x^4 r\theta \phi} = + \sqrt{\det g_{TN}}$.
Using this two form, a six-dimensional anti-self-dual three-form can be constructed as
\be
H^{(3)} = h(v)\, dv \wedge \omega_{TN},  \label{H-pert-BH}
\ee
where $ h(v)$ is an arbitrary function of $v$.  The metric sourced by such a form field is given by 
\be
ds^2 = \psi^{-1}(r) [du dv + (\psi(r) -1 + S(v, r)) dv^2] + \psi(r) \: ds^2_{TN},
\ee
with 
\be
S (v,r) = - \frac{4r}{R_4^2(4 r+R_4^2)}   h(v)^2. \label{S-TN}
\ee
The function $S (v,r)$ does not vanish at infinity, so once again the deformed metric does not look manifestly asymptotically flat. However, this can be readily fixed by shifting $u$ coordinate as \cite{Jatkar:2009yd},
\be
u \to u +  \frac{1}{R_4^2} \int^v_0   h(v')^2 dv'.
\ee
Once again, we can take the near-horizon limit of the deformed metric by taking the $r \rightarrow 0$ limit.

\mysection{Garfinkle-Vachaspati hair modes in the near-horizon region}
\label{sec:GV}

We start by considering Garfinkle-Vachaspati deformations in the near-horizon region \eqref{nearhorizon}.  As we reviewed in the previous section, the details of the Garfinkle-Vachaspati deformations are very different for the BMPV black hole in Taub-NUT space vs in flat space. The aim of this section to understand this difference from the near-horizon perspective.

The Garfinkle-Vachaspati deformation \eqref{GV_main} applied to the near-horizon metric  \eqref{nearhorizon} with $k = \partial_u$ results in,
\begin{equation}
ds^2= \frac{\ell^2}{z^2} \left(dudv  + dz^2 +  \left(\frac{z^2}{\ell^2} + H\right) dv^2 \right)+ \ell^2 d\Omega^2_{S^3}, \label{deformed_nearhorizon}
\end{equation}
where $H(v,z,\tilde{\theta},\tilde{\phi},\tilde{\psi})$ is a harmonic function for the six-dimensional metric \eqref{nearhorizon}. Expanding $H$ in terms of the $S^3$ spherical harmonics $H(v,z,\tilde{\theta},\tilde{\phi},\tilde{\psi}) = H^{I}(v,z)Y^{I}(\tilde\theta,\tilde \phi,  \tilde \psi)$ we get equations for functions $H^{I}(v,z)$,
\be 
z^3 \partial_z \left(\frac{1}{z}\partial_z H^{I}(v,z) \right) - L(L+2) H^{I}(v, z) =0, \label{gvequation}
\ee
where we have used the fact that the spherical harmonics of  $S^3$ satisfies $\nabla^2 Y^{I} = -L(L+2)Y^{I}$, with  $L$ taking values in
non-negative integers $\mathbb{N}$. Scalar spherical harmonics on  $S^3$ are labeled as $I = L,m_+ , m_-$ with $|m_\pm| \leq \frac{L}{2}$ and $\frac{L}{2} -m_{\pm} \in \mathbb{N}$. Explicitly, these spherical harmonics can be written as \cite{BenAchour:2015aah}
\be
Y^{I}(\tilde\theta,\tilde \phi,  \tilde \psi) \propto e^{i(S\tilde\phi + D\tilde\psi)}(1-x)^{\frac{S}{2}} (1+x)^{\frac{D}{2}} P^{(S,D)}_{\frac{L-S-D}{2}}(x),
\label{spherical_harmonics_general}
\ee
where $P^{(a,b)}_{n}$ are the Jacobi polynomials and $x=\cos2\tilde\theta,   S= m_+ + m_- ,   D= m_+ -m_- $.   We will see below that the black hole hair modes discussed in the previous section  correspond to different $L$ modes from the $AdS_3 \times S^3$ perspective.  In Poincar\'e coordinates $(\bar u, \bar v, \bar z)$  introduced in \eqref{Poincare}, the deformed metric \eqref{deformed_nearhorizon} takes the form
\begin{equation}
ds^2= \frac{\ell^2}{\bar z^2} \left(d \bar u \, d \bar v  + d\bar z^2 +  \bar H  d\bar v^2 \right)+ \ell^2 d\Omega^2_{S^3}, \label{deformed_nearhorizon_2}
\end{equation}
where the function $\bar H (\bar v, \bar z) $ satisfies the same equation as $H(v, z)$
\be 
\bar z^3 \partial_{\bar z} \left(\frac{1}{\bar z}\partial_{\bar z} \bar H^{I}(\bar v, \bar z) \right) - L(L+2) \bar H^{I}(\bar v, \bar z) =0\label{gvequation_2}.
\ee

\subsection{$L=0$ modes}
\label{sec:L_0}

With $L=0$, equation \eqref{gvequation_2} simply reduces to the  wave equation in  $AdS_3$.  Since there are no propagating degrees of freedom  in  three-dimensional $AdS_3$ gravity, we expect for \eqref{deformed_nearhorizon_2} either a trivial solution or  a solution which is related to $AdS_3 \times S^3$ by a coordinate transformation. Indeed, this expectation is realised. Solving equation \eqref{gvequation}, we get (the index $I$ takes the value $(0,0,0)$ and is  dropped for simplicity), 
\begin{equation}
  \bar H= a(\bar v) \bar z^2 + b(\bar v).  
\end{equation}

The term  corresponding to $b(\bar v)$ (non-normalisable mode) can be removed by transforming the coordinate $\bar u$,
\be
\bar u \to \bar u - \int^v b(\bar v') d \bar v'.
\ee
 The term  corresponding to $a(\bar v)$  (normalisable mode) can be removed by a transformation of the type discussed by Banados, Chamblin, and Gibbons \cite{Banados:1999tw},
\be
\bar v \rightarrow f(\bar v') \ , \qquad   \bar u \rightarrow \bar u' - \frac{\bar z'{}^2 f''(\bar v')}{2f'(\bar v')} \ , \qquad  \bar z \rightarrow \bar z'\sqrt{f'(\bar v')} \label{gcct1},
\ee
where $f'(\bar v'), f''(\bar v')$ are the first and the second derivatives of the function $f(\bar v')$ with respect to $\bar v'$. 
These transformations change the metric to
\begin{equation}
ds^2= \frac{\ell^2}{\bar z'{}^2} \left(d \bar u' d\bar v'  + d\bar z'{}^2\right) + \ell^2  d\Omega^2_{S^3},
\end{equation}
which is the original metric \eqref{nearhorizon_new} without the deformation in the new coordinates, provided the function $f(\bar v')$ is obtained via solving the equation,
\be
 a(f(\bar v')) f'(\bar v')^2   -  \frac{1}{2} S\{f(\bar v'),\bar v'\} =0, 
\ee
for the given function $a(\bar v)$. In this equation, $S\{f(x),x\}$ is the Schwarzian derivative of the function $f(x)$,
\be
S\{f(x),x\} = \frac{f'''(x)}{f'(x)} - \frac{3}{2} \left(\frac{f''(x)}{f'(x)}\right)^2.
\ee
\subsection{$L=1$ modes}

For  $L=1$ equation \eqref{gvequation} becomes,
\be
 z^3 \partial_z \left(\frac{1}{z}\partial_z H^{I}(z,v) \right) - 3 H^{I}(z,v) =0.
 \ee
 A general solution to this equation is given by 
 \begin{equation}
   H^{I}= c^{I}(v)z^3  + \frac{d^{I}(v)}{z}.
 \end{equation}
 The normalisable modes $c^{I}(v)z^3$ are not regular at the horizon $z \to \infty$. The non-normalisable modes $\frac{d^{I}(v)}{z}$ can be regular at the horizon.
 The index $I$ takes values $(1,m_+, m_-)$ with $m_+ = \pm \frac{1}{2} , m_- = \pm \frac{1}{2}$. Discarding the normalisable modes, we get, 
 \be
 H= \frac{1}{z} \sum_{m_+,m_-} \left( d^{1,m_+,m_-}(v) Y^{1,m_+,m_-}  \right). \label{H-flat}
\ee
 The spherical harmonics can be evaluated  as 
 \bea
 Y^{1, \pm\frac{1}{2},\pm \frac{1}{2}} &\propto& \sin\tilde\theta \ e^{\pm i \tilde \phi}, \\
 Y^{1, \pm\frac{1}{2},\mp \frac{1}{2}} &\propto& \cos\tilde\theta \ e^{\pm i\tilde \psi}.
 \eea 
 From \eqref{FLct} and \eqref{GH-r} we observe that $\frac{1}{z} \sim \tilde r$. Thus, we see that $H$ in equation \eqref{H-flat} correspond to $f_i (v)w^i$ with cartesian coordinates defined $w^i$ defined in \eqref{w-coords}. The form of the function $f_i (v)w^i$ is precisely the form that correspond to the Garfinkle-Vachaspati deformations for the BMPV black hole in flat space discussed in the previous section. Thus we conclude that  the non-normalisable $L=1$ modes in the near horizon region correspond to the Garfinkle-Vachaspati deformations for the BMPV black hole in flat space.

Is the deformed metric regular at the horizon $z\to \infty$?
In terms of the barred coordinates, the deformed near horizon metric takes the form,
\be
ds^2 = \frac{\ell^2}{\bar z^2}\left(d \bar u d \bar v + d\bar z^2 + \frac{\bar f_i (\bar v) n^i }{\bar z} d\bar v{}^2 \right) + \ell^2 d\Omega^2_{S^3}.
\ee
Here $n^i =w^i /|w|$ is a 4-dimensional unit vector. The horizon is at $\bar z\to \infty$. The functions $\bar f_i (\bar v)$ are related to the functions $ f_i (v)$ though they are not identical. With the coordinate transformation
\begin{align}
\bar v &= -\frac{1}{V} \  , & \  \bar z &=\frac{1}{VW} \ , & \   \bar u &= U + \frac{1}{VW^2},   \label{ctADS}
\end{align}
the deformed metric can be brought to the form,
\be
ds^2 = \ell^2 \left( W^2 dUdV + \frac{dW^2}{W^2} + \frac{W^3}{V} \bar f_i (\bar v(V)) n^i dV^2  \right) + \ell^2 d\Omega^2_{S^3}. \label{deformation_L_1}
\ee
The horizon is now located at $V = 0$. The metric appears singular at $V=0$. However, as earlier, we can ensure that the $g_{VV}$ component vanishes by a shift in the $U$ coordinate
\be
U = \widetilde U - W G(V,  \tilde \theta^i), \label{U-shift}
\ee
with
\be
G(V,  \tilde \theta^i) =    \int_0^V \frac{n^if_i(\bar v(V'))}{V'} dV' ,
\ee
Here $\tilde \theta^i$ collectively denotes $\tilde\theta, \tilde\phi,\tilde\psi$. Note that the function $G(V,  \tilde \theta^i) $ is such that in the limit $V \to 0$ it vanishes. 
Shift \eqref{U-shift} generates the following additional terms in the metric, 
\bea
& & - \ell^2 W^2  G(V, W, \tilde \theta^i) dWdV - \ell^2  W^3  \partial_{ \tilde \theta_i}  G(V, \tilde \theta^j) dW d \tilde \theta^i.
\eea 
These additional terms all vanish in the $V \to 0$ limit. The resulting metric is thus smooth near $V=0$, however, the $V$ derivatives of the metric are not. 
This leads to divergences in the Riemann tensor. For example, $R_{V W V W} $ diverges as $1/V$ as $V\to 0$. The conclusion is in line with the fact that these modes are also singular on the BMPV black hole horizon in the full black hole geometry \cite{Jatkar:2009yd}.

 \subsection{$L=2$ modes}
 
Next we consider $L=2$ modes.  For these modes, equation \eqref{gvequation} simplifies to 
\be
 z^3 \partial_z \left(\frac{1}{z}\partial_z H^{I}(z,v) \right) - 8 H^{I}(z,v) =0.
 \ee
The general solution to this equation is given by 
 \begin{equation}
   H^{I}= c^{I}(v)z^4  + \frac{d^{I}(v)}{z^2}.
 \end{equation}
 The black hole horizon is at $z\rightarrow \infty$ and hence we select the non-normalisable $\frac{d^{I}(v)}{z^2}$ part of the solution that seems regular at the horizon.\footnote{In a different context, the normalisable $c^{I}(v)z^4$ part of the solution was considered in \cite{Balasubramanian:2010ys}, though in a  different coordinate system. Ref.~\cite{Balasubramanian:2010ys} concluded that these modes are not regular at the horizon. The aim of ref.\cite{Balasubramanian:2010ys} was to  find solutions that maintain the self-dual orbifold asymptotics and hence they chose the normalisable modes. We are interested in the black hole hair modes. Therefore, to begin with we choose a solution that seems regular at the horizon.} The modes with specific values for the index $I$ can be identified with the Garfinkle-Vachaspati  hair modes for the black hole in Taub-NUT space. From \eqref{FLct} and \eqref{spherical_harmonics_general}, we observe that $x^4$ independent modes  correspond to $m_+ = 0$. The allowed values of $m_-$ are $m_- =  0,\pm 1$.  Thus, the general $x^4$ independent deformation takes the form,
 \be
 H= \frac{1}{z^2}\left(d^{(2,0,0)}(v)Y^{(2,0,0)} + d^{(2,0,-1)}Y^{(2,0,-1)} + d^{(2,0,1)}Y^{(2,0,1)}  \right)\label{spherharmTN}.
 \ee
 These spherical harmonics can be evaluated  as 
 \begin{align}
& Y^{(2,0,0)}    \propto \cos \theta, &
&Y^{(2,0,1)}  \propto \sin \theta \,  e^{i \phi}, &  
& Y^{(2,0,-1)} \propto \sin \theta \, e^{-i \phi}.&
 \end{align} 
 We see that the function $H$ in \eqref{spherharmTN} corresponds to $g_i (v)y^i$  for $i=1,2,3$ and hence to the Garfinkle-Vachaspati  hair modes for the black hole in Taub-NUT space \eqref{gvTN}. Indeed, by the taking the $r \to 0$ limit of the GV deformed black hole metric in Taub-NUT we obtain $AdS_3 \times S^3$ metric \eqref{deformed_nearhorizon} with $H$ given as in \eqref{spherharmTN}.

In $(U, V, W)$ coordinates, cf.~\eqref{ctADS}, the deformed metric takes the form
\be
ds^2 = \ell^2 \left( W^2 dUdV + \frac{dW^2}{W^2} + W^4 g_i (\bar v(V)) n^i dV^2 \right) + \ell^2 d\Omega^2_{S^3}. \label{deformation_L_2}
\ee
The functions $\bar g_i(\bar v)$ is related to the functions $ g_i (v)$ though they are not identical.  A key difference compared to the $L=1$ deformed metric \eqref{deformation_L_1} is that the coefficient of $dV^2$ terms now goes as $W^4$. The metric appears regular at $V =0$. Though, there is a catch: as $V \to 0$, $\bar v$ coordinate changes rapidly from a finite value to infinity. Thus, it is not obvious in these coordinates if metric \eqref{deformation_L_2} is regular or not at $V=0$. To address this, we can ensure that $g_{VV}$ vanishes by a shift in the $U$ coordinate
\be
U = \widetilde U - W^2 G(V,  \tilde \theta^i),
\ee
with
\be
G(V, \tilde \theta^i) =  \int_0^V n^ig_i(v(V')) dV' .
\ee
The shift generates additional terms
\bea
& & - 2 \ell^2 W^3    G(V, \tilde \theta^i) dWdV - \ell^2  W^4  \partial_{\tilde \theta^i}  G(V, \tilde \theta^i) dW d \tilde \theta^i.
\eea
These additional terms all vanish in the $V \to 0$ limit. The resulting metric is therefore smooth near $V=0$, however, $V$ derivatives of the metric are not. Specifically,  $\partial_V^2 G$ diverges at $V=0$. These divergences, however, do not appear in the Riemann tensor components.  

We conclude that the three functions $g_i(v)$ generate smooth deformations of the  near-horizon geometry.\footnote{We can now qualify the statement  in \cite{Balasubramanian:2010ys} that  $L=2$ modes are not regular in the bulk. We can pick $L=2$ solutions which are, in fact, regular in the bulk but are non-normalisable.} This conclusion is in line with the analysis of \cite{Jatkar:2009yd} where it  was shown that the three Garfinkle-Vachaspati  hair modes for the black hole in Taub-NUT space are also regular at the horizon.

\mysection{Form field hair modes in the near horizon region}
\label{sec:form_field}

We now consider a deformation of $AdS_3 \times S^3$ due to the anti-self-dual form field $H_{MNP}$.  The relevant perturbation equations were given in \cite{Maldacena:1998bw}. For our analysis the most relevant equation is the 
anti-self-duality condition for the perturbing form field with mixed components. 
This equation takes the form \cite[eq.~(4.18)]{Maldacena:1998bw} 
\be
H_{ab\mu} + \frac{1}{2}\epsilon_{\mu}{}^{\nu\rho}\epsilon_{ab}{}^{c}H_{c\nu\rho} =0,
\label{asd_condition}
 \ee
 where latin indices  $a,b,c, \dots$ stand for the sphere coordinates and greek indices $\mu, \nu, \rho, \dots$ for the AdS coordinates. For all six-dimensions we use $M, N, P, \dots$ indices. Let $B_{MN}$ be the two-form for which $H_{MNP}$ is the field strength, i.e., $H_{MNP} = \partial_M B_{NP} + \partial_{N} B_{PM}  + \partial_P B_{MN}.$

 Consider expanding $B_{MN}$ in spherical harmonics. To identify the relevant black hole hair modes, we do not need the most general decomposition. We only need the decomposition of $B_{MN}$ with one sphere coordinate and one AdS coordinate. 
The components $B_{\mu a}$ can be expanded in terms of spherical harmonics on $S^3$  as
\be
B_{\mu a}= \sum b^{I}_\mu Y^{I}_a,  
\label{B_simp_2}
\ee
where $Y^{I}_a$ are the one-form harmonics. It turns out, we do not even need general properties of one-form harmonics. As it will become shortly, we only need one of the simplest one-form harmonic obtained by lowering the Killing vector \cite{BenAchour:2015aah}
\be
\partial_{\tilde \phi} + \partial_{\tilde \psi}.
\ee
In components this one-form is
\be
\xi = \cos^2 \tilde \theta d \tilde \phi + \sin^2 \tilde \theta d\tilde \psi.
\ee
This one-form satisfies 
\be
\bar{\epsilon}_{a}{}^{bc} \partial_b \xi_c = - 2 \xi_a,
\ee
where $\bar{\epsilon}_{abc}$ is the epsilon tensor for the unit-sphere. We use the convention $\bar {\epsilon}_{\tilde \theta  \tilde \phi \tilde \psi} = \sin \tilde \theta \cos \tilde \theta$.
Eq.~\eqref{B_simp_2} now simplifies to
\be
B_{\mu a} = b_\mu \xi_a.
\label{B_simp}
\ee

Computing $H_{MNP}$ for the $B$-field of the form \eqref{B_simp} and substituting it in \eqref{asd_condition}, we get
 \be
 \frac{1}{2}\epsilon_{\mu}{}^{\nu\rho}(\partial_\nu b_\rho -\partial_\rho b_\nu)= \frac{1}{\ell}\eta b_\mu \label{bfield1},
 \ee
 with $\eta = -2$. For later use we keep $\eta$ general. To solve equation \eqref{bfield1} we make the ansatz that  only the $b_v$ component of the one-form $b_\mu$ is non-vanishing.\footnote{Applying $\nabla^\mu$ on \eqref{bfield1} we see that $ \nabla^\mu b_\mu =0$, where $\nabla^\mu$ is the covariant derivative compatible with the three-dimensional $AdS_3$ part of the metric. We can use this condition to set $b_z = 0$. Next, requiring $db$ to be compatible with the $k^M$ Killing symmetry, we have $\pounds_k db = d (i_k db) =0$. Choosing $i_k db =0$ gives us $i_k (\star_3 b) = 0$ from \eqref{bfield1}, which implies $\star_3 (b \wedge k) = 0$. This condition can be used to set $b_u=0$. } Furthermore, we assume that the perturbation is compatible with $\partial_u$ Killing symmetry, i.e., $b_v$ only depends on the $v$ and $z$ coordinates.  Substituting this ansatz in equation \eqref{bfield1}, we get
 \begin{equation}
   z\partial_z b_v (v,z) = \eta b_v (v,z),  \qquad \text{where} \qquad \epsilon_{uvz}= \frac{\ell^3}{2z^3}.
 \end{equation} 
 This gives as a solution
 \begin{equation}
    b_v = c(v)z^{\eta},
 \end{equation}
 which in turn gives, 
 \begin{align}
B_{v \tilde \phi} &= \frac{c(v) }{ z^2}  \cos^2 \tilde \theta , &
B_{v \tilde \psi} &= \frac{c(v)}{z^2}  \sin^2 \tilde \theta . \label{B_mu_a_simple}
\end{align}
As will shortly become clear, this perturbation describes the anti-self-dual form-field perturbations to the black hole in Taub-NUT space from the near horizon region perspective. 

To see this, let us recall that the three-form hair perturbation for the  black hole in Taub-NUT space takes the form \eqref{H-pert-BH}.
 In the near horizon limit $r\to 0$ the $H$-field becomes,
\be
H = -  \frac{h(v) r}{R_4^2} \sin \theta \, dv \wedge d\theta \wedge d \phi + \frac{h(v) }{R_4^2} dv \wedge dr \wedge (dx^4 + \cos \theta d\phi), 
\ee
 which in coordinates $(u,v,z, \tilde \theta, \tilde \phi, \tilde \psi)$, takes the form
\begin{align}
H_{v \tilde \theta \tilde \phi} &= - \frac{\ell^4}{R_4^2 z^2} h(v) \sin \tilde \theta \cos \tilde \theta, &
H_{v \tilde \theta \tilde \psi} &= \frac{\ell^4}{R_4^2 z^2} h(v) \sin \tilde \theta \cos \tilde \theta, &  \label{H1} \\
H_{v z \tilde \phi} &= - \frac{\ell^4}{R_4^2 z^3} h(v)  \cos^2 \tilde \theta, &
H_{v z \tilde \psi} &=- \frac{\ell^4}{R_4^2 z^3} h(v)  \sin^2 \tilde \theta. \label{H2}
\end{align}
The $B_{MN}$ sourcing this $H_{MNP}$ can be chosen to be of the form 
\eqref{B_mu_a_simple} with $c(v) = -\frac{\ell^4}{2R_4^2}  h(v)$.

 Now that we have a solution to the anti-self dual form-field, we can consider its  back reaction on the metric and solve the corresponding Einstein equation with the stress tensor sourced by the form-field deformation.  In our conventions, Einstein equation with anti-self-dual form field as an additional source take the form \cite{Jatkar:2009yd, Chakrabarti:2020ugm}
 \be
 R_{MN} = F_{MPQ}F_{N}{}^{PQ} + H_{MPQ}H_{N}{}^{PQ}. \label{sd-asd}
 \ee

 With the  $H_{MNP}$ given in \eqref{H1}--\eqref{H2} a simple calculation shows that it only sources the $vv$ component of the Einstein equation, and moreover the sphere dependence drops out. We find\footnote{It is simplest to write this result in mixed component. We will see shortly that the left hand side of the Einsteins equations take  a particularly simple form with mixed indices of this type.}
  \begin{equation}
 H^{uPQ}H_{vPQ} = \frac{8 \ell^2}{R_4^4} \left(\frac{h(v)^2}{z^2}\right). 
 \end{equation}
To capture the deformation in the metric we make the ansatz, 
\begin{equation}
ds^2= \frac{\ell^2}{z^2} \left(dudv  + dz^2 +  \left(\frac{z^2}{\ell^2} + S(v,z)\right) dv^2 \right)+ \ell^2 d\Omega^2_{S^3}, \label{deformed_S}
\end{equation}
where we have introduced the function $S(v, z)$ in the $dv^2$ term. Only  $ R^{u}{}_{v}$ component of the  Einstein equation gives a non-trivial equation. We find, 
\be
z^3 \partial_z \left(\frac{1}{z}\partial_z S(v,z) \right) +   \frac{8 \ell^4}{R_4^4} \left(\frac{h(v)^2}{z^2}\right) =0.
\ee
This equation can be readily solved for $S$ to yield,
\be
S(v, z) = - \frac{\ell^4}{R_4^4} \frac{h(v)^2}{z^2}.
\ee
This is precisely the $r\to0$ limit of the function $S$ that appears in equation \eqref{S-TN}.  We conclude that  the B-field 
\eqref{B_mu_a_simple} together with metric \eqref{deformed_S}
describes the anti-self-dual form-field perturbations to the black hole in Taub-NUT space from the near horizon region perspective.

We also note that the three-dimensional part of the metric \eqref{deformed_S} is of the following special form, 
\begin{equation}
 g_{\mu\nu} = g_{\mu\nu}^{AdS_3} - 4 \ell^2 R_{4}^{-4} h(v)^2 k_\mu k_\nu, \label{null-warped-metric}
\end{equation}
where $k = \partial_u$.

In $(U, V, W)$ coordinates, cf.~\eqref{ctADS}, the deformed metric takes the form
\be
ds^2 = \ell^2 \left( W^2 dUdV + \frac{dW^2}{W^2} - W^4 \bar h(\bar v(V))^2 dV^2 \right) + \ell^2 d\Omega^2_{S^3}. \label{metric_deformation_S}
\ee
The function $\bar h(\bar v)$ is related to the function $ h (v)$ though they are not identical. 
As far as the powers of $V$ and $W$ are concerned, this metric has the same form as \eqref{deformation_L_2}. Therefore, the arguments given below \eqref{deformation_L_2} regarding the smoothness of the deformation apply without any change. The form field $H_{MNP}$ is also regular in  $(U, V, W)$ coordinates. We conclude that the anti-self-dual form-field gives rise to a  smooth deformation of the  near-horizon geometry. This conclusion is in line with the analysis of \cite{Jatkar:2009yd} where it  was shown that the anti-self-dual form-field  hair modes for the black hole in Taub-NUT space are regular at the horizon.  

\mysection{Fermionic hair modes in the near horizon region}
\label{sec:fermionic}
In this section, we  make some observations on the non-normalisability of the fermionic hair modes of refs.~\cite{Jatkar:2009yd, Banerjee:2009uk, Chakrabarti:2020ugm} in  $AdS_3 \times S^3$. We find it easiest to present this analysis in six-dimensions.  The relevant linearised equations for the gravitino $\psi^\alpha_M$ are 
\be
\Gamma^{MNP}D_N\Psi^\alpha_P- {F}^{MNP}\Gamma_N{\widehat{\Gamma}}^1_{\alpha \beta}\Psi^\beta_P=0. \label{Gravitinoeqn} 
\ee
Our aim is to solve this equation in the near horizon geometry. We work with the form of the near horizon geometry \eqref{nearhorizon_GH}.  Making the ansatz,
$ \Psi^\alpha_M=0,$ for $M\neq v,$ and $\partial_u \Psi^\alpha_v=0$, 
and choosing  the gauge condition
$ 
\Gamma^v \Psi_v^\alpha=0,
$
we obtain a class of solutions
\be
   \Psi_v \propto r^{3/2}\varepsilon(v,\theta,\phi) \quad \text{for} \quad {\widehat{\Gamma}}^1\varepsilon=-\varepsilon,
 \ee
 with the angular dependence given as
 \begin{align}
 	& \varepsilon(v,\theta,\phi)= h(v)e^{i\phi/2}\begin{pmatrix}
 		\cos (\theta/2)\\
 		-\sin (\theta/2)
 	\end{pmatrix}, \quad \text{or} & 
 	&  \varepsilon(v,\theta,\phi)= h(v)e^{-i\phi/2}\begin{pmatrix}
 		\sin (\theta/2)\\
 		\cos (\theta/2)
 	\end{pmatrix} .&
 \end{align}
 The same solutions are obtained by taking the $r \to 0$ limit of the corresponding expressions in \cite{Jatkar:2009yd, Chakrabarti:2020ugm}. These solutions are regular at the horizon.\footnote{We do not present  a smoothness analysis in this paper. However, it is clear that a straightforward adaptation of the analysis of ref.~\cite{Jatkar:2009yd, Chakrabarti:2020ugm} is applicable. } 
A key point that we wish to highlight is that in $(u, v, z)$ coordinates, cf.~\eqref{FLct},
the gravitino deformation grows near the boundary $z=0$ of $AdS_3$ as
\be
 \Psi_v \propto \frac{1}{z^3}\varepsilon(v,\tilde{\theta},\tilde{\phi},\tilde{\psi}).
\ee 
The gravitino hair modes are non-normalizable in the near-horizon region. 
     
\mysection{Discussion}
\label{sec:discussion}

In this paper, we discussed the hair modes of the BMPV black hole in flat space and in Taub-NUT space from the near-horizon $AdS_3 \times S^3$ perspective.  We found that the non-trivial hair modes involve non-zero angular momentum on the $S^3$ and are regular at the $AdS_3$ horizon. Most importantly, all non-trivial hair modes change the asymptotics of the near-horizon geometry.  Our results clearly show that these modes ``leak-out'' from near-horizon region to the asymptotically flat region: they grow towards the AdS boundary before falling off to zero in the asymptotically flat region.

We wish to make a few technical remarks regarding the above analysis. 
\begin{enumerate}

\item Using the AdS/CFT dictionary, we can identify the dimensions of the CFT operators that correspond to the bulk deformations discussed above. Turning on non-normalisable modes  correspond to deformations of the CFT with operators of  conformal dimension $\Delta > 2$. For operators dual to scalars of mass $m$ in $AdS_3$, conformal dimension $\Delta$ is
\be
\Delta = 1 + \sqrt{1 + m^2 \ell^2}.
\ee 
The Garfinkle-Vachaspati deformations in $AdS_3 \times S^3$ with $L=1$ and $L=2$ can be identified with scalars of masses $m^2 \ell^2 = L(L+2)$. Hence, the smooth hair mode with $L=2$ corresponds to a deformation of the CFT by an operator of mass dimensions  
\be
\Delta = 1 + \sqrt{1 + 8} = 4.
\ee The same conclusion can be reached by considering the Garfinkle-Vachaspati deformation as a metric deformation. Since the deformation term grows as $z^{-4}$ near the boundary $z=0$, it corresponds to a deformation of the  CFT by an operator of mass dimensions  
$\Delta =4$, see, e.g., comments in \cite{Herdeiro:2002ft}.

\item We can similarly identify the dimensions of the operators that correspond to the form-field deformation.
Taking one more exterior derivative of equation \eqref{bfield1}, we can convert it into the equation for a massive vector field
\begin{equation}
  \star_3 d \star_3 d b  = \ell^{-2} \eta^2 b. \label{massive-vector}
\end{equation} 
 A deformation by a massive vector field of mass $m^2 \ell^2 =\eta^2 =4$ corresponds to deforming the boundary CFT by an operator of  dimension $\Delta = 1 + \sqrt{m^2 \ell^2} = 3$ \cite{Herdeiro:2002ft}.  In addition, since the form-field back reacts and generates a non-normalisable metric deformation that grows as $z^{-4}$ near the boundary $z=0$, we conclude that the CFT deformation is accompanied with an additional deformation by a $\Delta = 4$ operator.

\item In section \ref{sec:form_field}, we made the observation that the three-dimensional part of the metric \eqref{deformed_S} can be  written  in a special form \eqref{null-warped-metric}. For the case when the function $h(v)$ is  a constant, we recognise metric \eqref{null-warped-metric} as the null warped $AdS_3$ metric (and also as Schr\"odinger metric with $z=2$).  As is well known, these metrics can be obtained as solutions to three-dimensional Einstein gravity with negative cosmological constant coupled to a massive vector field \cite{Guica:2011ia}. As discussed in the previous remark, the metric deformation discussed in section \ref{sec:form_field} can also be thought of as a deformation sourced by a massive vector field \eqref{massive-vector}. Thus, the fact that the metric \eqref{deformed_S} can be  written  in the null warped $AdS_3$ form is perhaps not unexpected. Null warped $AdS_3$ metrics are known to be duals of deformations of the boundary CFT by operators of dimension $4$ \cite{Guica:2010sw}. This is in line with the observations made above. It will  be interesting to identify such operators in the D1-D5 CFT and  explore the connection to the null warped $AdS_3$ further. 
 
\end{enumerate}

In refs.~\cite{Mathur:2011gz, Lunin:2012gp,  Mathur:2012tj} various deformations of $AdS_3 \times S^3$ were considered  in the context of the fuzzball program. Even though the starting point is the same, there are many differences between our results and their analyses. Firstly, since there are no horizons in the fuzzball cases, these references did not need to consider the regularity at the horizon. The analyses in these references is instead in  global AdS.   Secondly,  the metric perturbations considered in these references do not have direct analogs in our analysis, but roughly speaking a class of modes considered in \cite{Mathur:2011gz} corresponds to our $L=0$ Garfinkle-Vachaspati modes.\footnote{See also point 3 on page 26 of ref.~\cite{Mathur:2011gz} where a related observation is made.}

 In ref.~\cite{Virmani:2022glo}, it was suggested that fuzzballs should also admit hair modes like the ones that we have considered in this paper. Specifically, by putting fuzzballs in Taub-NUT space, it was conjectured that there should be fuzzball hair, like the ones obtained by putting the BMPV black hole in  Taub-NUT space. We hope to report our progress on constructing such hair on a class of simplest fuzzballs in the future.

\subsection*{Acknowledgments} SB and YKS thank CMI Chennai for warm hospitality. The work of AV was partly supported by SERB Core Research Grant CRG/2023/000545 and by the ``Scholar in Residence'' program of IIT Gandhinagar. 
\appendix

\mysection{Hair deformations using six dimensional supergravity}
\label{app:6D_sugra}
The set-up we are interested in is six-dimensional minimal supergravity coupled to one tensor multiplet. The bosonic equations of motion for this set-up are given as \cite{Cariglia:2004kk}
\bea
 R_{MN} &=& 2\nabla_{M} \phi \nabla_{N}\phi+ e^{2\sqrt{2}\phi} \left(G_{MPQ}G_{N}{}^{PQ}-\frac{1}{6}g_{MN}G_{PQR}G^{PQR}\right), \label{A.1}\\
dG&=&0, \\
d \star_6 (Ge^{2\sqrt{2}\phi})&=&0,\\
\nabla^2 \phi &=& \frac{1}{3\sqrt{2}}e^{2\sqrt{2}\phi}G_{PQR}G^{PQR}. \label{A.4}
\eea
For the constant dilaton case $G^2=G_{PQR}G^{PQR}$ must vanish
and we have 
\bea
dG &=&0, \\
d \star_6 G&=&0, \\ 
R_{MN} &=& G_{MPQ}G_{N}{}^{PQ}.
\eea 
Breaking $G_{MNP}$ into self-dual $F_{MNP}$ and anti self-dual $H_{MNP}$ parts, we get  equation \eqref{sd-asd}.

   Bena, Giusto, Shigemori, and Warner 
    \cite{Bena:2011dd}
   wrote a linear system of equations for supersymmetric solutions to the above set-up \eqref{A.1}--\eqref{A.4}. The metric ansatz considered in their work is of the form
\begin{equation}
ds^2=-2H^{-1}(dv+\beta)\left(du+\omega+ \frac{1}{2}{\cal F}(dv+\beta)\right)+H ds_4^2,
\label{metric_main}
\end{equation}
where $H$ and ${\cal F}$ are functions and  $\beta$, $\omega$ are one-forms on the four dimensional base with metric $ds_4^2 = h_{mn} dx^m dx^n$. A priori, the functions $H$, ${\cal F}$, the one-forms $\beta$, $\omega$, and the metric $h_{mn}$ all depend on $v, x^m$ but not on $u$. We consider the case when the base metric is hyper-K\"ahler. Moreover, we restrict to the cases where the one-form $\beta$ and the base metric are taken to be independent of the $v$ coordinate. Under these assumptions, the exterior derivative of the one-form $\beta$ becomes self-dual on the base space 
\be
\star_4 \, \widetilde{d}\beta = \widetilde{d} \beta,
\ee
where $\widetilde{d}$ is the exterior derivative restricted to the base space. Ultimately, we are interested in the constant dilaton case, however, to make connection with     \cite{Bena:2011dd} it is better to keep $\phi$ non-vanishing for now. Furthermore, we will be only interested in the cases when the  hyper-K\"ahler base is either flat space or Taub-NUT space. These assumptions already make the linear system of  Bena, Giusto, Shigemori and Warner manageable. For the cases we are interested in a further drastic simplification occurs, namely
\be
\beta = 0. 
\ee
This is sometimes called  a ``trivial fibration'' in the literature.
This makes the linear system very manageable.

Under these assumptions, the tensor gauge field $G_{MNP}$ takes the form,
\bea
G &=& d \left[-\frac{1}{2}Z_1^{-1}(du+\omega)\wedge dv \right]+ \widehat{G}_1,\\
e^{2\sqrt{2}\phi} \star_6 G &=&  d \left[ -\frac{1}{2} Z_2^{-1}(du+\omega)\wedge dv \right]+ \widehat{G}_2,
\eea
where\footnote{Our Hodge star conventions are the same as \cite{Bena:2011dd}.}
\bea
\widehat{G}_1 &=& \frac{1}{2} \star_4 (DZ_2)+ dv \wedge\Theta_1,\\
\widehat{G}_2 &=& \frac{1}{2} \star_4 (DZ_1)+ dv \wedge\Theta_2,
\eea
and where for our simplified set-up $D\Phi=\widetilde{d}\Phi - \beta \wedge \dot{\Phi} = \widetilde{d}\Phi$. 
$\Theta_1$ and $\Theta_2$ are two two-forms on the base space. $Z_1$ and $Z_2$ are two functions on the base space. Physically speaking,  $Z_1$ and $Z_2$ are the electric potentials and $\Theta_1$ and $\Theta_2$ are the magnetic two-forms. The BPS conditions imply that the electric potentials $Z_1$ and $Z_2$ and related to the function $H$ that appears in the metric \eqref{metric_main} as
\begin{align}
 Z_1&=H e^{\sqrt{2}\phi}, &
 Z_2&= H e^{-\sqrt{2}\phi}.
\end{align}
For the hyper-K\"ahler base space cases, the magnetic two-forms on the base are self-dual 
\begin{align}
\star_4 \Theta_1 &= \Theta_1, &
\star_4 \Theta_2 &= \Theta_2. \label{self-duality-Theta}
\end{align}

The ``first layer'' of the BPS equations determines $(Z_1, Z_2)$ and $(\Theta_1,\Theta_2)$. The equations  are,
\begin{align}
 D \star_4 D Z_1 &= 0, &
 D \star_4 D Z_2 &= 0, \label{Z1-Z2-equations}
\end{align}
and
\begin{align}
\widetilde{d}\Theta_2 &=\frac{1}{2}\star_{4}D\dot Z_1, &          \widetilde{d}\Theta_1&=\frac{1}{2}\star_4D\dot Z_2 \label{Theta-Z-eqs}
\end{align}
where  over-dots  are the Lie derivative with respect to $v$.

The ``second layer'' of the BPS equations determines the function ${\cal F}$ and the one-form $\omega$. The one-form $\omega$ carries information about the angular momentum of the solution and the function ${\cal F}$ about the left-moving momentum.  The equations are,
\bea        
  D\omega+ \star_4D\omega &=& 2Z_1\Theta_1+ 2Z_2\Theta_2, \label{omega-eq} \\
     \star_4 D \star_4 \left(\dot{\omega}-\frac{1}{2}D{\cal F}\right)&=& -2 \star_4\Theta_1\wedge \Theta_2+ Z_1\partial_v^2 Z_2+ Z_2\partial_v^2 Z_1+ \partial_v Z_1\partial_v Z_2. \label{F-eq}
\eea
 
 \subsubsection*{Underlying black hole: BMPV}
 We can embed $J_L=0$ BMPV black hole in this formalism rather easily. The black holes correspond to no $v$-dependence and no magnetic fluxes $\Theta_1 = \Theta_2 = 0$.  Furthermore, for our set-up, the dilaton $\phi$ is set to its constant asymptotic value. We can adjust this value such that  $\phi = 0$, i.e., $Z_1=Z_2=H$. For the non-rotating black hole the one form $\omega =0$. Equation \eqref{Z1-Z2-equations} then implies $H$ is harmonic; equation \eqref{F-eq} tells us that ${\cal F}$ is harmonic. In Gibbons-Hawking coordinates \eqref{GH-flat} the choice $H = \psi(r)$ and ${\cal F} = \psi(r) -1$, cf.~\eqref{psi-r},  gives us the $J_L=0$ BMPV black hole. We can check that $2G=d(\psi^{-1}du\wedge dv)+\star_4DH$ matches with $F^{(3)}=r_0(\star_6\epsilon_3+ \epsilon_3)$, cf.~\eqref{F3}, with $\epsilon_3= \sin\theta dx^4\wedge d\theta\wedge d\phi$.

The embedding of $J_L \neq 0$ BMPV black hole additionally requires introducing a $v$-independent one form $\omega$ such that
\be
  D\omega+ \star_4D\omega =0. \label{omega-simple}
\ee
As a result \eqref{omega-eq} is satisfied.
The functions ${\cal F}$ and $H$ remain the same. The one-form $\omega$ enters the three-form $G$. It is straightforward to see that the one-form $\zeta$ of \cite{Chakrabarti:2020ugm} satisfies \eqref{omega-simple} and gives the correct three-form $G$.

   \subsubsection*{Underlying black hole: BMPV in Taub-NUT}
We take the base metric $ds_4^2=ds_{TN}^2$, cf.~\eqref{metric-TN}.
As in the previous paragraph we take $\phi =0$, i.e., $Z_1=Z_2=H$. Equation \eqref{Z1-Z2-equations} implies $H$ is harmonic on the Taub-NUT base. Equation \eqref{F-eq} tells us that ${\cal F}$ is harmonic on the Taub-NUT base.  In Gibbons-Hawking coordinates \eqref{metric-TN} the choice $H = \psi(r)$ and ${\cal F} = \psi(r) -1$, cf.~\eqref{psi-r},  gives us the $J_L=0$ BMPV black hole in Taub-NUT. We can check that $2G=d(\psi^{-1}du\wedge dv)+\star_4DH$ matches with $F^{(3)}=r_0(\star_6\epsilon_3+ \epsilon_3)$, cf.~\eqref{F3_TN}, with $\epsilon_3= \sin\theta dx^4\wedge d\theta\wedge d\phi$. The embedding of the $J_L \neq 0$ BMPV black hole in Taub-NUT is exactly the same as discussed above except that $\star_4$ in \eqref{omega-simple} is now with respect to the Taub-NUT metric \eqref{metric-TN}.

           \subsubsection*{Garfinkle-Vachaspati hair modes}
           In both the above cases, equation \eqref{F-eq} tells us that ${\cal F}$ is harmonic on the base space. Thus,  apart from $\psi-1$ we have the freedom of adding a harmonic function to ${\cal F}$. This freedom is precisely the same as applying the Garfinkle-Vachaspati transform.

\subsubsection*{Form-field hair modes}
 To  turn on the anti-self-dual field deformations on the above discussed black hole in Taub-NUT, we take in addition
 \begin{equation}
 \Theta_1=h(v) \omega_{TN}=-\Theta_2, 
\end{equation}
where $\omega_{TN}$  is introduced in \eqref{omega_TN}. Since 
$\omega_{TN}$ is self-dual, equations \eqref{self-duality-Theta} are satisfied. Since $\omega_{TN}$ is closed, i.e., $d \omega_{TN} = \widetilde{d} \omega_{TN}  =0$, equations \eqref{Theta-Z-eqs} imply that we can continue to choose $H$ to be $v$-independent harmonic function. Then, equation \eqref{omega-eq} reduces to
     \begin{equation}
      D\omega+ \star_4D\omega=d\omega+ \star_4d\omega=2Z_1\Theta_1+2Z_2\Theta_2=0.  
     \end{equation}
Thus, $\omega=0$ continues to be the solution for the non-rotation case. For rotating case    $\omega=\zeta$ continues to be the solution.  The only non-trivial change is in $\mathcal{F}$.  Equation \eqref{F-eq} simplifies to
\begin{equation}
 \star_4 d \star_4 d \mathcal{F}=4  \star_4 \Theta_1\wedge\Theta_2=-4h(v)^2  \star_4 \omega_{TN}\wedge\omega_{TN}.
 \end{equation}
 Solution to this equation for ${\cal F}$ is $S(v,r)$ as given in \eqref{S-TN}.

\end{document}